\documentclass[12pt]{article}
\usepackage{cite}
\usepackage{epsfig}
\usepackage{amsfonts,amsmath}
\usepackage{hhline}
\usepackage{amssymb}
\usepackage{times}

\newlength{\dinwidth}
\newlength{\dinmargin}
\newcommand{\ycut}{\mbox{$y_{\rm cut}$}}
\newcommand{\GeV}{\mbox{\rm GeV}}
\newcommand{\GeVsq}{\mbox{${\rm GeV}^2$}}
\newcommand{\etbreit}{\mbox{$E_{T, {\rm Breit}}$}}
\newcommand{\etabreit}{\mbox{$\eta_{\mbox{\scriptsize Breit}}$}}
\newcommand{\rcone}{\mbox{$R_0$}}

\newcommand{\nsub}{\mbox{$\langle N_{\rm subjet}(y_{\rm cut}) \rangle$}}

\begin{document}

\begin{titlepage}
\noindent
DESY 98--210  \hfill  ISSN 0418--9833 \\
December 1998 \\

\begin{center}

\mbox{}
\vspace{1.2cm}

\begin{Large}
{\bf \Large
Measurement of Internal Jet Structure in \\
Dijet Production in Deep-Inelastic Scattering at HERA\\}
\vspace{2.2cm}
H1 Collaboration
\end{Large}
\end{center}
\vspace{1.3cm}
\begin{abstract}
\noindent
Internal jet structure in dijet production in deep-inelastic scattering
is measured with the H1 detector at HERA.
Jets with transverse energies $\etbreit > 5\,\GeV$ are
selected in the Breit frame employing $k_\perp$ and cone jet algorithms.
In the kinematic region of squared momentum transfers 
$10 < Q^2 \lesssim 120\, \GeVsq$ ~and $x$-Bjorken values  
$2 \cdot 10^{-4} \lesssim x_{\rm Bj} \lesssim 8 \cdot 10^{-3}$,
jet shapes and subjet multiplicities are measured as a function of 
a resolution parameter.
Distributions of both observables are corrected for detector effects 
and presented as functions of the transverse jet energy and jet 
pseudo-rapidity.
Dependences of the jet shape and the average number of subjets
on the transverse energy and the pseudo-rapidity of the jet are observed.
With increasing transverse jet energies and decreasing pseudo-rapidities,
i.e.\ towards the photon hemisphere, the jets are more collimated.
QCD models give a fair description of the data.
\end{abstract}

\vspace{2cm}
\begin{center} \sl  submitted to Nuclear Physics B
\end{center}

\end{titlepage}

\begin{center} \Large  H1 Collaboration
\end{center}
\noindent
 C.~Adloff$^{34}$,                
 V.~Andreev$^{25}$,               
 B.~Andrieu$^{28}$,               
 V.~Arkadov$^{35}$,               
 A.~Astvatsatourov$^{35}$,        
 I.~Ayyaz$^{29}$,                 
\linebreak
 A.~Babaev$^{24}$,                
 J.~B\"ahr$^{35}$,                
 P.~Baranov$^{25}$,               
 E.~Barrelet$^{29}$
% ...

\bigskip
\noindent
% ...
$ ^{24}$ Institute for Theoretical and Experimental Physics,
          Moscow, Russia \\
 $ ^{25}$ Lebedev Physical Institute, Moscow, Russia$^{f,k}$ \\
% ...
$ ^{28}$ LPNHE, \'{E}cole Polytechnique, IN2P3-CNRS, Palaiseau, France \\
 $ ^{29}$ LPNHE, Universit\'{e}s Paris VI and VII, IN2P3-CNRS,
          Paris, France \\
% ...
$ ^{34}$ Fachbereich Physik, Bergische Universit\"at Gesamthochschule
          Wuppertal, Wuppertal, Germany$^a$ \\
 $ ^{35}$ DESY, Institut f\"ur Hochenergiephysik, Zeuthen, Germany$^a$ \\
% ...
$ ^a$ Supported by the Bundesministerium f\"ur Bildung, Wissenschaft,
        Forschung und Technologie, FRG,
        under contract numbers 7AC17P, 7AC47P, 7DO55P, 7HH17I, 7HH27P,
        7HD17P, 7HD27P, 7KI17I, 6MP17I and 7WT87P \\
% ...
$ ^f$ Supported by the Deutsche Forschungsgemeinschaft \\
% ...
$ ^k$ Supported by Russian Foundation for Basic Research 
       grant no. 96-02-00019

\newpage
 
\section{Introduction}
A sizeable fraction of the final states produced in high
energy collisions shows the characteristic feature of
large amounts of hadronic energy in small angular regions.
These collimated sprays of hadrons (called jets) are the observable 
signals of underlying short distance processes 
and are considered
to be the footprints of the underlying partonic final states.
Quantitative studies of jet production require a precise jet
definition, which is given by a jet finding algorithm.

Jets so defined exhibit an internal structure which is sensitive
to the mechanism by which a complex aggregate of observable hadrons 
evolves from a hard process. 
The understanding of this mechanism involves 
higher orders of the strong coupling constant in
perturbation theory as well as non-perturbative contributions.
This is a challenging task for theory.
Recently, for some specific hadronic final state quantities,
encouraging results have been obtained by exploiting the characteristic 
power behaviour of non-perturbative effects 
and by analytical, approximate calculations of perturbative QCD
parton evolution down to the semi-soft regime~\cite{evshape,ochs}.
Furthermore, since jet production rates
are used to test the predictions of perturbative QCD, the understanding
of their detailed properties and internal structure is 
an important prerequisite.

The internal structure of jets has been studied in $e^+ e^-$\cite{eejet}
and in hadron-hadron collisions~\cite{ppjet}.
At the $e^\pm p$ collider HERA, these investigations can be performed in
photoproduction ($Q^2 \approx 0 \; \GeVsq$) and in deep-inelastic scattering (DIS)
at large squared four momentum transfers $Q^2$.
In a previous publication we have measured the 
$E_T$ dependence of the jet width~\cite{h1gammap}
in photoproduction.
Recently, the ZEUS collaboration has investigated jet shapes 
in photoproduction~\cite{zeusphoto}
and in DIS at $Q^2 > 100$~\GeVsq~\cite{zeusdis}.
Both analyses are carried out in the laboratory frame. 
This means that for DIS at high $Q^2$ mostly events
with only one jet enter the analysis.

The hadronization of the current jet in deep-inelastic scattering 
in the Breit frame has already been studied with event shape 
variables~\cite{h1evtshape}, charged particle multiplicities
and fragmentation functions~\cite{h1multiplicity}.
In this paper we take the first steps towards a complete
understanding of jet properties in DIS.
We analyse the hadronization of jets 
in multijet production in the Breit frame.
The Breit frame, where the virtual photon interacts head-on with the 
proton, has been chosen in this analysis
because here the produced transverse\footnote{transverse with
respect to the $z$-axis which is given by the axis of the virtual 
photon and the proton.}
energy, \etbreit, directly reflects the hardness of the 
underlying QCD process.
We present measurements of internal jet structure
in a sample of inclusive dijet events with transverse jet energies of
$\etbreit > 5$~\GeV, $10 < Q^2 \lesssim 120\,\GeVsq$ and 
$2 \cdot 10^{-4} \lesssim x_{\rm Bj} \lesssim 8 \cdot 10^{-3}$.
This is the \etbreit ~range where jet cross section measurements are 
currently performed at HERA and compared to perturbative QCD 
calculations (e.g.~\cite{h1r2cone,h1loqdijet}).
The analysis is based on data taken in 1994 with the H1 detector
at HERA when $27.5$\,\GeV 
~positrons collided with $820$\, \GeV ~protons. 
The data correspond to an integrated luminosity of 
${\cal L}_{\rm int} \simeq 2\,\mbox{pb}^{-1}$.

Jets are defined in the Breit frame by $k_\perp$ and cone jet algorithms.
Two observables, jet shapes and, for the first time,
subjet multiplicities, are studied.
The jet shape measures the radial distribution of the 
transverse jet energy around the jet axis.
For the $k_\perp$ cluster algorithm we have also measured the
multiplicity of subjets, resolved at a resolution scale which is 
a fraction of the jet's transverse energy.
Both observables are presented for different ranges of the transverse 
jet energy and the pseudo-rapidity\footnote{The pseudo-rapidity $\eta$ 
is defined as $\eta \equiv - \ln ( \tan \theta / 2 )$ where $\theta$ is 
the polar angle with respect to the proton direction. 
This definition is chosen
in both the laboratory frame and the Breit frame.}
of the jets in the Breit frame.

The paper is organized as follows.
Section \ref{detector} gives a brief description of the H1 detector.
In section \ref{jetdef} we introduce the jet algorithms used in the
analysis and give the definition of the measured observables in section 
\ref{observables}.
In section \ref{models} we give a short description of the QCD models 
which are used for the correction of the data and to which the results are later
compared (in section \ref{compare}).
The data selection and the correction procedure are described in sections
\ref{dataselection} and \ref{correction} and the results 
are discussed in section \ref{results}.

\section{The H1 Detector \label{detector}}
A detailed description of the H1 detector can be found elsewhere~\cite{H1det}.
Here we briefly introduce the detector components relevant 
for this analysis: the liquid argon (LAr) calorimeter~\cite{larcalo},
the backward lead-scintillator calorimeter (BEMC)~\cite{bemc},
and the tracking chamber system~\cite{tracker}.

The hadronic energy flow is mainly measured by the
LAr calorimeter extending over the polar angular range
$4.4^\circ < \theta <  154^\circ$ with full azimuthal coverage.
The polar angle $\theta$ is defined with respect to the proton
beam direction ($+z$ axis).
The LAr calorimeter consists of an electromagnetic section ($20-30$ 
radiation lengths) with lead absorbers and a hadronic section with 
steel absorbers.
The total depth of both calorimeters 
varies between $4.5$ and $8$ interaction lengths. 
Test beam measurements of the LAr~calorimeter modules show an
energy resolution of 
$\sigma_{E}/E\approx 0.50/\sqrt{E\;[\GeV]} \oplus 0.02$  for 
charged pions~\cite{pions}.
The absolute scale of the hadronic energy is known for the present 
data sample to $4\%$.

The scattered positron is detected by the BEMC 
with a depth of $22.5$ radiation lengths covering the 
backward region of the detector,
$155^\circ < \theta < 176^\circ$.
The electromagnetic energy scale is known to an accuracy of $1\%$.

The calorimeters are 
surrounded by a superconducting solenoid providing a uniform
magnetic field of $1.15$ T parallel to the beam axis in the tracking region.

Charged particle 
tracks are measured in two concentric jet drift chamber modules
(CJC), covering the polar angular range $ 15^\circ < \theta < 165^\circ$.
The forward tracking detector covers $7^\circ < \theta < 25^\circ $ 
and consists of drift chambers with alternating planes 
of parallel wires and others with wires in the radial direction.
A backward proportional chamber (BPC) with an angular
acceptance of $151^\circ < \theta < 174.5^\circ$ improves the 
identification of the scattered positron.
The spatial resolution for reconstructed BPC hits is about 1.5\,mm in the
plane perpendicular to the beam axis.

\section{Jet Definitions \label{jetdef}}
The jet algorithms used in this analysis are applied 
to the particles boosted into the Breit frame.
Particle refers here either to an energy deposit in the detector
(see section \ref{dataselection}), to a stable hadron
or a parton in a QCD model calculation.
In all cases the scattered positron is excluded.
The Breit frame is defined by $\vec{q} + 2 x_{\rm Bj} \vec{P} = 0$,
where $\vec{q}$ and $\vec{P}$ are the momenta of the exchanged boson
and the incoming proton. 
The $z$-axis is defined as the direction of the incoming proton.

In the following analysis we use two different jet definitions:
a cone algorithm and a $k_\perp$ cluster algorithm.
Both jet definitions are invariant under boosts along the $z$-direction.
The recombination of particles is carried out in the $E_T$ recombination 
scheme, which is based on transverse energies $E_T$, pseudo-rapidities $\eta$ 
and azimuthal angles $\phi$ of the particles. 
The transverse energy and the direction of a jet are defined by
\begin{equation}
E_{T, {\rm jet}} = \sum_{i} E_{T, i} , \hskip8mm
\eta_{\rm jet} = \frac{\sum_i E_{T,i} \; \eta_i}{\sum_{i} E_{T,i}} , \hskip8mm
\phi_{\rm jet} = \frac{\sum_i E_{T,i} \; \phi_i}{\sum_{i} E_{T,i}} ,
\label{recomb}
\end{equation}
where the sums run over all particles $i$ assigned to the jet\footnote{All
particles are considered massless by setting $E_i = |\vec{p_i}|$.}.

\subsection{Cone Algorithm}
Based on the original proposal of Sterman and 
Weinberg~\cite{sterman} many different
implementations of cone algorithms have been developed.
While the basic idea of the cone algorithm is simple and very intuitive,
an operational definition is non-trivial.
The resulting jet cross sections depend on how the algorithm
treats the choice of jet initiators and configurations of overlapping 
jet cones.
It has repeatedly been pointed out that many definitions of 
cone algorithms are not infrared and/or 
collinear safe~\cite{soper,seymour}.

In this analysis we use the definition implemented in the 
algorithm PXCONE~\cite{pozo}
which does not suffer from the problems discussed in ~\cite{soper,seymour}.
This definition, which corresponds closely to the Snowmass 
proposal~\cite{snowmass} and to the algorithm used in the 
CDF experiment~\cite{cdfcone}, is also used by the 
OPAL collaboration~\cite{opal}.

Particles are assigned to jets based on their spatial distance $R$
in pseudo-rapidity and azimuth space
($R^2 =\Delta \eta^2 + \Delta \phi^2$).
The algorithm operates as follows:

\begin{enumerate}

\item
Each particle is considered 
as a seed of a jet, for which steps 2-4 are performed.

\item
The jet momentum is calculated from all particles within a cone of
radius \rcone ~around the seed direction using eq.\
(\ref{recomb}).

\item
If the jet direction differs from the seed direction, the jet 
direction is taken as the new seed direction and step 2 is repeated. 

\item
When the jet direction is stable
the jet is stored in the list of ``protojets"
(if it is not identical with a protojet already found).

\item
The steps 2 to 4 are repeated for all midpoints of pairs of protojets 
as seed directions\footnote{In practice it is sufficient 
to do this only for pairs of protojets with a distance between 
\rcone ~and $2\,\rcone$.}. 
This leads to the infrared safety of the procedure~\cite{seymour}.

\item 
Protojets with transverse energies of $E_{T,{\rm jet}} < \epsilon$ are removed 
from  the list. 
The cut-off parameter $\epsilon$ specifies below which transverse
energies protojets are not considered in the overlap treatment (steps 7-8).

\item
All remaining protojets that have more than a fraction $f$ 
of their transverse energy contained in a protojet of higher transverse 
energy are deleted.

\item
All particles that are contained in more than one protojet are
assigned to the protojet whose center is nearest in $(\eta,\phi)$.

\item 
The jet momenta are recalculated using eq.\ (\ref{recomb}).
All protojets with $E_{T, {\rm jet}} < \epsilon$ are deleted and the remaining 
ones are called jets.

\end{enumerate}

The jets with the highest transverse energies are considered in the analysis.
Due to the reassignment of particles to jets and the recalculation of the 
jet axis (steps 7, 8) it may happen that single particles within a jet 
have a distance larger than \rcone ~to the jet axis.
This analysis is made with the parameter settings 
$\epsilon = 5\,\GeV$, $f = 0.75$ and a cone radius of $R_0 =1.0$.

\subsection{\boldmath Inclusive $k_\perp$ Algorithm}
The ambiguities that occur for cone jet definitions
(choice of seeds, overlapping cones) are 
avoided in cluster algorithms which successively recombine
particles to jets.
One definition of such an algorithm (proposed in~\cite{invkt} and
implemented in the KTCLUS algorithm~\cite{ktclus}) 
has properties very similar to cone algorithms.
As in the cone algorithm the clustering procedure is based 
on the longitudinally boost-invariant quantities 
$E_T, \Delta \eta, \Delta \phi$.
The minimum of all distances between particles
is determined and either the corresponding pairs of particles are 
merged into pseudo-particles or single (pseudo-) particles are declared as jets.
This process is iterated until no particles are left:

\begin{enumerate}
\item
We start with a list of all particles and an empty list of jets.

\item
For each particle $i$ as well as for each pair of particles ($i,j$)
the distances $d_i$ and $d_{ij}$ are calculated
\begin{equation}
d_i = E_{T,i}^2 \; R_0^2
\hskip5mm \mbox{and}  \hskip5mm 
d_{ij} = \min (E_{T,i}^2 , E_{T,j}^2) \; R^2_{ij}
\hskip5mm {\rm with } \hskip5mm 
 R^2_{ij} = \Delta \eta_{ij}^2 + \Delta \phi_{ij}^2 \, .
\end{equation}

\item 
The smallest value of all the $d_i$ and $d_{ij}$ is labeled 
$d_{\rm min}.$

\item
If $d_{\rm min}$ belongs to the set of 
$d_{ij}$, the particles $i$ and $j$
are merged into a new particle using the recombination prescription
in eq.\ (\ref{recomb}) and removed from the list of particles.

\item
If $d_{\rm min}$ belongs to the set of $d_{i}$, 
the particle $i$ is removed from the
list of particles and added to the list of jets.

\item
When no particles are left (i.e.\ all particles are included in jets)
the procedure is finished. 

\end{enumerate}

The last jets that entered the list are the ones with highest 
transverse energies. 
These jets are considered in the analysis.
This jet definition implies that particles with $R_{ij} < R_0$ are subsequently 
merged, so that all final jets are separated by distances $R_{ij} > R_0$.
It is still possible that particles inside a jet have a distance
$R_{ij} > R_0$ to the jet axis and that particles with
$R_{ij} < R_0$ are not part of the jet.
The parameter $\rcone$ is set to $\rcone = 1.0$.

\section{The Observables \label{observables}}
Two observables of internal jet structure are investigated in this 
analysis. They are sensitive to different aspects of jet broadening.

The jet shapes are studied for the cone and the $k_\perp$ algorithm.
This observable measures the radial distribution of the 
transverse jet energy only and is affected by hard and by soft 
processes over the whole radial range.

A natural choice for studying the internal structure of jets with the
$k_\perp$ cluster algorithm is the multiplicity of subjets, resolved at 
a resolution scale which is a fraction of the jet's transverse energy.
These subjet multiplicities are sensitive to more local structures 
of relative transverse momentum within a jet.
Here the perturbative and the non-perturbative contributions are
better separated.
While at larger values of the resolution parameter
perturbative contributions dominate, at smaller values 
non-perturbative contributions become increasingly important.

\subsection{The Jet Shape}
The jet shape $\Psi(r)$ is defined as the fractional transverse jet
energy contained in a subcone of radius $r$ concentric with the jet axis, 
averaged over all considered jets in the event sample
\begin{equation}
\Psi(r) \equiv \frac{1}{N_{\rm jets}} \sum_{\rm jets} \;
\frac{E_{T}(r)}{E_{T, {\rm jet}}} \, ,
\label{eq:defpsi}
\end{equation}
where $N_{\rm jets}$ is the total number of these jets.
As proposed in~\cite{seymour},
only particles assigned by the jet algorithm to the jet are considered.

Usually the denominator in the definition of $\Psi$ is given by the
summed $E_T$ of all particles within a radius \rcone ~to the jet 
axis. 
This means that $\Psi(r/\rcone = 1) = 1$.
In our definition (\ref{eq:defpsi}) of $\Psi$ the denominator is given by the
transverse energy of the jet. 
Since neither for the cone nor for the $k_\perp$ definition are all particles
necessarily assigned to a jet within a radius of $r/\rcone<1$ to the jet 
axis, $\Psi(r/\rcone = 1)$ is not constrained to have the value of one.
With this choice of our observable we are also sensitive to 
the amount of transverse jet energy outside the radius \rcone.

\subsection{Subjet Multiplicities}
For each jet in the sample the clustering procedure is repeated for 
all particles assigned to the jet.
The clustering is stopped when the distances $y_{ij}$ between
all particles $i,j$ are above some cut-off \ycut
\begin{equation}
y_{ij}  \; =  \;\frac{\min (E_{T,i}^2 , E_{T,j}^2)}{E_{T, {\rm  jet}}^2} \; \;
\frac{\left( \Delta \eta_{ij}^2 + \Delta \phi_{ij}^2 \right)}{R_0^2}
 \; >  \; \ycut 
\end{equation}
and the remaining (pseudo-)particles are called subjets.
The parameter \ycut ~defines the minimal relative transverse
energy between subjets inside the jet and thus determines the extent 
to which the internal jet structure is resolved.
From this definition it follows that for 
$\ycut > 0.25$ no subjet is resolved
(therefore the number of subjets is one), 
while for $\ycut \to 0$ every particle
in the jet is a subjet.
The observable that is studied in this analysis is the average number
of subjets for a given value of the resolution parameter,
for values $\ycut \ge 10^{-3}$.

\section{QCD Models \label{models}}
A simulation of the detailed properties of the hadronic
final state is available in the form of Monte Carlo
event generators. They include the matrix
element of the hard subprocess in first order of the
strong coupling constant $\alpha_s$, approximations of
higher order QCD radiation effects, and a model to describe the
non-perturbative transition from partons to hadrons.

The LEPTO Monte Carlo~\cite{lepto} incorporates the
${\cal O} (\alpha_s)$ QCD matrix element 
and takes higher order parton emissions 
to all orders in $\alpha_s$ approximately into account 
using the concept of 
parton showers~\cite{shower} based on the 
leading logarithm DGLAP equations~\cite{dglap}.
QCD radiation can occur before and after
the hard subprocess. The formation of hadrons 
is performed using the LUND string model~\cite{lund} 
implemented in JETSET~\cite{jetset}.

The HERWIG Monte Carlo~\cite{herwig} also includes
the  $\cal{O}$$(\alpha_s)$ QCD matrix element, but
uses another implementation of the parton shower cascade
which takes coherence effects fully into account.
The hadronization is simulated with the cluster fragmentation model~\cite{cluster}.

In ARIADNE~\cite{ariadne} 
gluon emissions are treated by the colour dipole model~\cite{cdm}
assuming a chain of independently radiating
dipoles spanned by colour connected partons. The first emission
in the cascade is corrected to reproduce the matrix element to first 
order in $\alpha_s$~\cite{ariadneme}. 
 
DJANGO~\cite{django} provides an interface between the event generators
LEPTO or ARIADNE and HERACLES~\cite{heracles} which makes it possible to include
${\cal O}(\alpha)$ QED corrections at the lepton line.

\section{Data Selection \label{dataselection}} 
The analysis is based on H1 data taken in $1994$ corresponding to
an integrated luminosity of ${\cal L}_{int} \simeq 2\,\mbox{pb}^{-1}$.
The event selection closely follows that described in a previous 
publication~\cite{h1r2cone}.
DIS events are selected where the scattered positron is measured
in the acceptance region of the BEMC at energies where trigger
efficiencies are approximately 100\,\%.
To ensure a good identification of the scattered positron
and to suppress background from misidentified 
photoproduction events the following cuts
are applied:
\begin{itemize}

\item
The cluster of the positron candidate must have an
energy-weighted mean transverse radius below $5\,\mbox{cm}$.

\item
A reconstructed BPC hit within $5\,\mbox{\rm cm}$ of the straight line
connecting the shower center with the event vertex is required.

\item
The $z$ position of the reconstructed event vertex must be 
within $\pm 30\,\mbox{\rm cm}$ of the nominal position.

\item
A cut on $35\,\GeV < \sum (E - p_z) < 70\,\GeV$ is applied,
where the sum runs over all energy deposits in the calorimeter.
In neutral current DIS events without undetected photon radiation the 
quantity $\sum (E - p_z)$ is expected to be equal to twice the energy 
of the initial state positron.
This cut reduces the contribution from photoproduction events as well 
as events where hard photons are radiated collinear to the incoming 
positron.
\end{itemize}

The event kinematics are calculated from the
polar angle $\theta_{el}$ and the energy $E'_{el}$ of the scattered positron
via $Q^2_{el} = 2\,E_0\,E'_{el}\,(1+\cos{\theta_{el}})$,
$y_{el}=1- E'_{el}/(2\,E_0) (1-\cos{\theta_{el}})$ and
$x_{\rm Bj} = Q^2 / (sy)$.
$E_0$ denotes the energy of the incoming positron
and $s$ the $ep$ centre-of-mass energy squared.
Events are only accepted, if $E'_{el} > 11$~\GeV,
$156^\circ < \theta_{el} < 173^\circ$, $Q^2 > 10\,\GeVsq $
and $y > 0.15$.
The resulting kinematic range is $10 < Q^2 \lesssim 120\,{\rm GeV}^2$
and $2 \cdot 10^{-4} \lesssim x_{\rm Bj} \lesssim 8 \cdot 10^{-3}$.

Jets are defined by the algorithms described in section \ref{jetdef}.
The input for the jet algorithms consists of a combination of 
energy clusters from the calorimeter and track momenta
measured in the central and forward trackers 
(as described in \cite{h1r2cone}).
While all energy clusters are considered, the four momentum of 
each single
track is only allowed to contribute up to a momentum of $350\,\mbox{\rm MeV}$.
This procedure partly compensates for energy losses in the
calorimeter due to dead material and noise thresholds.
It reduces the dependence of the jet finding efficiency on
the pseudo-rapidity of the jet and improves the reconstruction of the 
transverse jet energy~\cite{clim}.

The objects from tracking and calorimeter information are boosted to 
the Breit frame where the jet algorithms are applied.
We select events with at least two identified jets with transverse energies
of $\etbreit > 5\,\GeV$ in 
$-1 < \eta_{\mbox{\scriptsize jet,lab}} <  2$.
The two jets with the highest \etbreit ~are considered in the analysis.
The event sample for the inclusive $k_\perp$
algorithm (the cone algorithm) consists of 2045 (2657) dijet events.

\section{Correction of the Data \label{correction}}
The data are corrected for detector effects and
QED radiation from the lepton.
The detector response is determined using events from
Monte Carlo event generators that were subjected to a detailed 
simulation of the H1 detector. 
The following event generators are used: ARIADNE interfaced in DJANGO
(with and without the inclusion of QED corrections) 
and LEPTO.
Both generators give a good description of the kinematic variables 
of the inclusive DIS data sample as well as of the angular and transverse 
energy distributions of the jets~\cite{sonne}. 
We also observe a reasonable description of the observables introduced 
in section \ref{observables} (see section \ref{compare}).

The measured data points are corrected bin-by-bin for detector effects.
Using the generated event samples,
the correction factor for each bin is determined 
as the ratio of the generated value of the observable
and the value that is reconstructed after detector simulation.
These correction factors are independent of the inclusion of
QED radiation effects as included in DJANGO.
Their dependence on details of the modeling of the hadronic
final state is taken into account by considering the difference
between the correction factors from ARIADNE and LEPTO as 
systematic uncertainty.

For the $k_\perp$ (cone) algorithm the corrections for $\Psi(r)$ are below 
$10\,\%$ ($13\,\%$) for subcone radii $r>0.3$ and always below $27\,\%$ 
($23\,\%$).
The corrections for \nsub ~are always below $7\%$.
The correction factors from both QCD models are in good agreement
(they differ typically by not more than $2\,\%$)
for the jet shapes as well as for the subjet multiplicities~\cite{sonne}. 
The final correction factors are taken to be the mean values of the 
two models, taking the spread as the error.
In addition we have varied the calibration of the hadronic energy scale 
in the data sample in the range of $\pm 4\%$ around the nominal value. 
The error is estimated as the maximal deviation from the results
at the nominal value. 
For all observables it is at most $5\%$.
The overall systematic error is calculated by adding the 
errors from the model dependence and from the uncertainty
of the hadronic energy scale in quadrature.
In all figures the statistical and systematic errors
are added in quadrature.
Since each jet enters in all bins of a distribution, all errors
are correlated.

The background from misidentified photoproduction events is estimated 
with a sample of photoproduction events generated with PHOJET~\cite{phojet}
and is found to be negligible.

\section{Results \label{results}}
The jet shape and the subjet multiplicity are presented 
as functions of quantities directly related to the single jets,
namely the transverse jet energy (\etbreit) and the
pseudo-rapidity (\etabreit) in the Breit frame.
We also investigated whether the observables depend on the 
event kinematics.
The jet shapes and subjet multiplicities were compared for two bins
of $Q^2$ ($Q^2 < 20\,{\rm GeV}^2$ and $Q^2 > 20\,{\rm GeV}^2$)
and $x_{\rm Bj}$ 
($x_{\rm Bj} < 8 \cdot 10^{-4}$ and $x_{\rm Bj} > 8 \cdot 10^{-4}$)
respectively.
No dependence on $Q^2$ and $x_{\rm Bj}$ has been observed.

\subsection{Jet Shapes}
\begin{figure}
\begin{center}
\epsfig{file=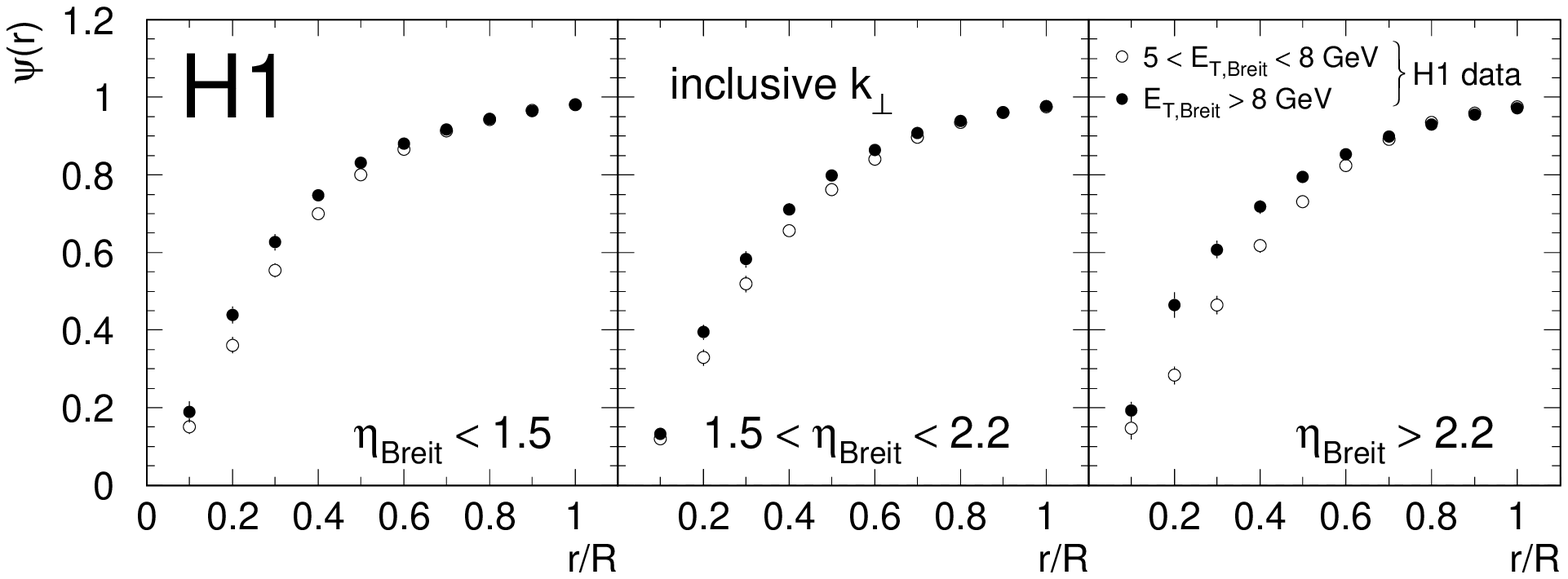,width=15.5cm}
\end{center}
\caption{The dependence of the jet shapes on the transverse jet energy
in three pseudo-rapidity regions.
The jet shapes at higher and at lower transverse jet 
energies for the inclusive $k_\perp$ algorithm are overlaid.
The comparison is shown as a function of the jet pseudo-rapidity in 
the Breit frame (positive pseudo-rapidities are towards the proton direction).
\label{fig:kt_shape_et}}
\end{figure}

\begin{figure}
\begin{center}
\epsfig{file=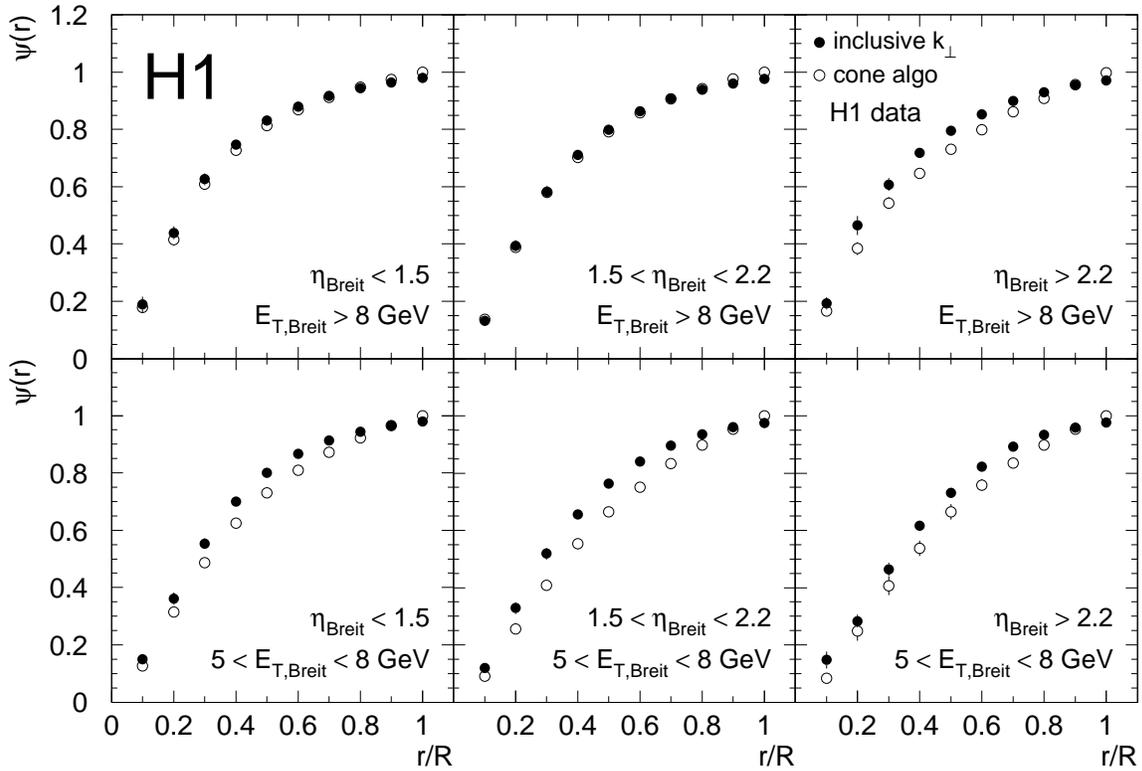,width=15.5cm}
\end{center}
\caption{Comparison of the jet shapes for the 
inclusive $k_\perp$ algorithm and the cone algorithm.
The data are shown as a function of the transverse jet energy and the jet
pseudo-rapidity in the Breit frame (positive pseudo-rapidities
are towards the proton direction).
\label{fig:kt_cone_shape}}
\end{figure}

The radial dependence of the jet shape $\Psi(r)$ for 
the $k_\perp$ algorithm is shown in Fig.~\ref{fig:kt_shape_et} 
in different ranges of the pseudo-rapidity in the Breit frame.
The results for jets of transverse energies
$5 < \etbreit < 8\,{\rm GeV}$ and $\etbreit > 8\,{\rm GeV}$ 
are superimposed.
The jet shape $\Psi(r)$ increases faster with $r$ for jets
at larger transverse energies, indicating that these jets are more collimated.
The same tendency is seen for the jets defined by the cone algorithm
which are compared to the jets found by the $k_\perp$ algorithm
in Fig.~\ref{fig:kt_cone_shape}.
For both jet definitions we also observe a dependence of the jet shape
on the pseudo-rapidity of the jets.
Jets towards the proton direction (at larger values of \etabreit)
are broader than jets towards the photon direction 
(smaller \etabreit).
In the region where the jets are most collimated
($\etbreit > 8\,{\rm GeV}$ and $\etabreit < 2.2$), 
very similar jet shapes are observed for the $k_\perp$ and 
cone algorithms.
The broadening of the jets for smaller \etbreit ~and larger 
\etabreit ~is more pronounced for the cone jet definition.

Recently jet shapes have been measured in dijet production 
in photon-photon collisions~\cite{ggshape} for jets defined 
by a cone algorithm at transverse energies comparable to 
those presented here.
The jet shapes in photon-photon collisions 
(where no $\eta$ dependence is observed)
are very similar to those measured in DIS in the 
Breit frame at $\etabreit < 1.5$.

\subsection{Subjet Multiplicities}
\begin{figure}
\begin{center}
\epsfig{file=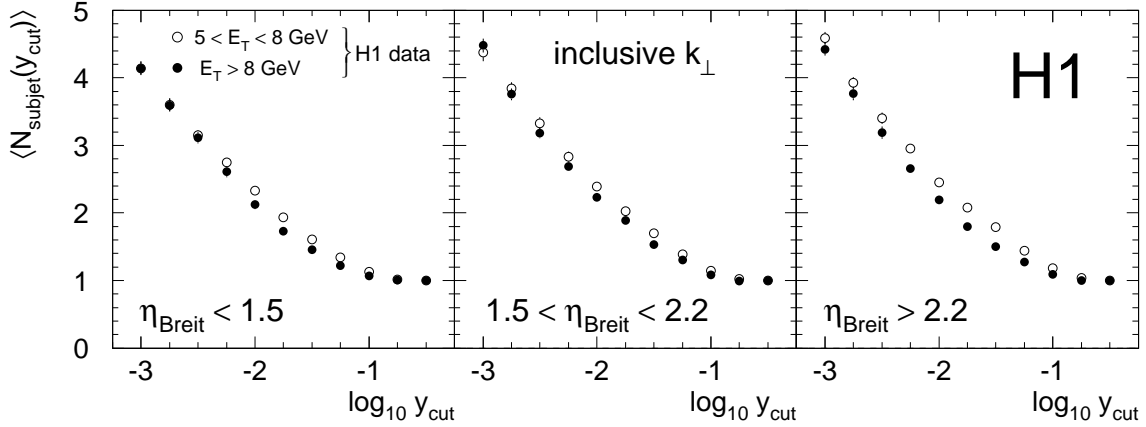,width=15.5cm}
\end{center}
\caption{The average number of subjets as a function of
the resolution parameter \ycut ~for the inclusive $k_\perp$ algorithm. 
The data are shown in $\etabreit$ ~bins for different $\etbreit$.
\label{fig:kt_subjet_et}}
\end{figure}

The subjet multiplicities for the $k_\perp$ algorithm 
are displayed in Fig.~\ref{fig:kt_subjet_et}.
The average number of subjets \nsub ~as a function
of the subjet resolution parameter at $\ycut \ge 10^{-3}$
is plotted.
Towards smaller values of \ycut, an increasing number of jet 
fragments with smaller relative transverse momenta is resolved.
The number of subjets at a given value of $y_{\rm cut}$
reflects the amount of relative transverse momentum with respect 
to the jet axis.
The subjet multiplicity is therefore a measure of the broadness of the jet.

At $y_{\rm cut} = 10^{-3}$ a jet is on average resolved into
$4.1$ -- $4.6$ subjets, depending on \etbreit ~and \etabreit ~of the 
jet\footnote{On average the jets in the data (as in the
simulated events) consist of eleven calorimetric energy clusters. 
For the LEPTO generator this is also approximately the 
average multiplicity of stable particles inside the jets.}.
For almost all values of $y_{\rm cut}$ the subjet multiplicity is 
larger for jets at smaller \etbreit ~and larger \etabreit,
indicating broader jets.

\begin{figure}
\begin{center}
\epsfig{file=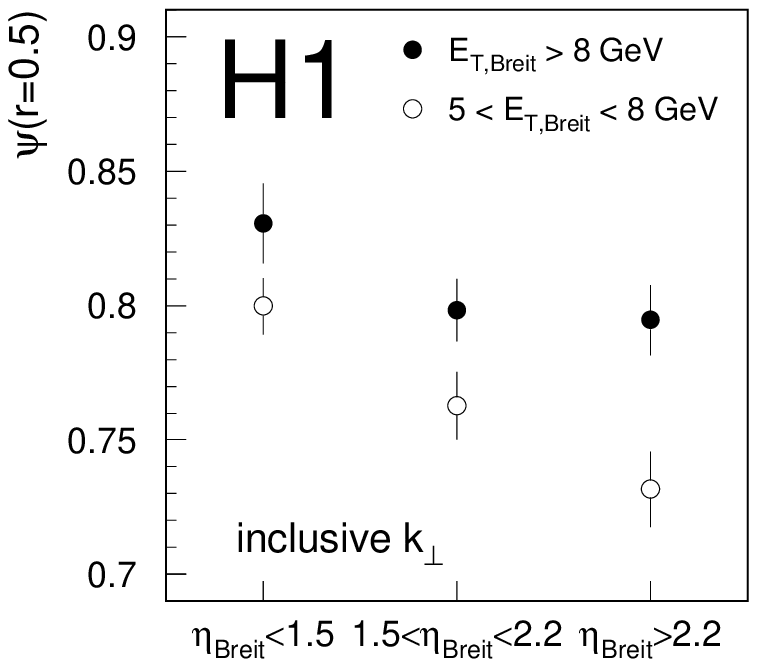,width=7.1cm}
\hskip8mm
\epsfig{file=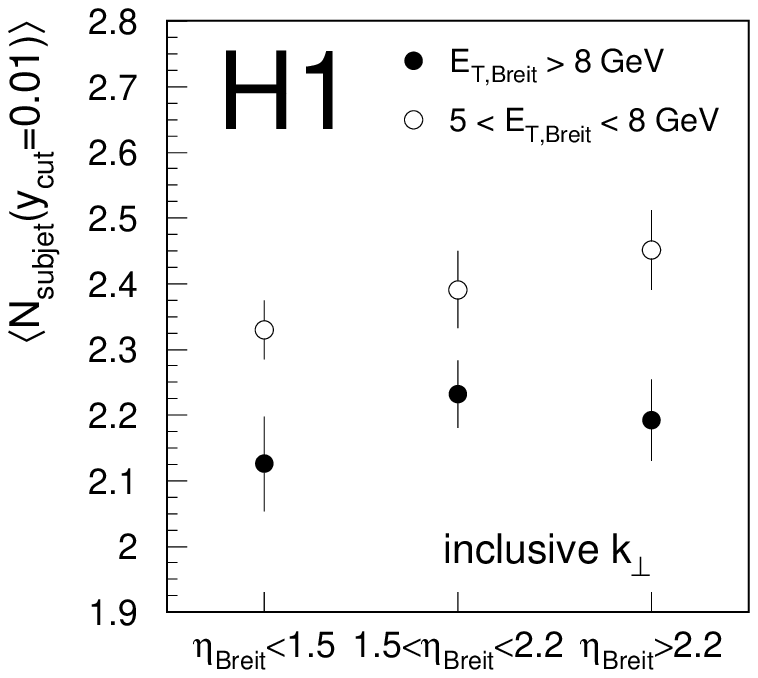,width=7.1cm}
\end{center}
\caption{
The jet shape $\Psi(r)$ at a fixed subcone radius $r=0.5$
in three \etabreit ~ranges and two \etbreit ~bins for the
inclusive $k_\perp$ algorithm (left).
The average subjet multiplicity at a fixed resolution parameter
$\ycut = 0.01$ 
in three \etabreit ~ranges and two \etbreit ~bins
for the inclusive $k_\perp$ algorithm (right).
Positive pseudo-rapidities are towards the proton direction. 
\label{fig:summary}}
\end{figure}

A summary of the results for both observables is given 
in Fig.~\ref{fig:summary}.
Here the \etbreit ~and the \etabreit ~dependence of the
jet shape and the average number of subjets are
shown at an intermediate value of the resolution parameter
(jet shape: $r=0.5$ and subjet multiplicity: $y_{\rm cut} = 10^{-2}$).

Although the subjet multiplicities are sensitive to the jet 
broadening in a different way  than the jet shapes, consistent
conclusions can be drawn for both measurements.
The jet broadening depends on both the transverse jet energy as well
as the pseudo-rapidity in the Breit frame.
While the pseudo-rapidity dependence is most pronounced at smaller 
transverse jet energy, the transverse energy dependence is stronger
in the forward region (at larger pseudo-rapidities).

\section{Comparison with QCD Model Predictions \label{compare}}
\begin{figure}
\begin{center}
\epsfig{file=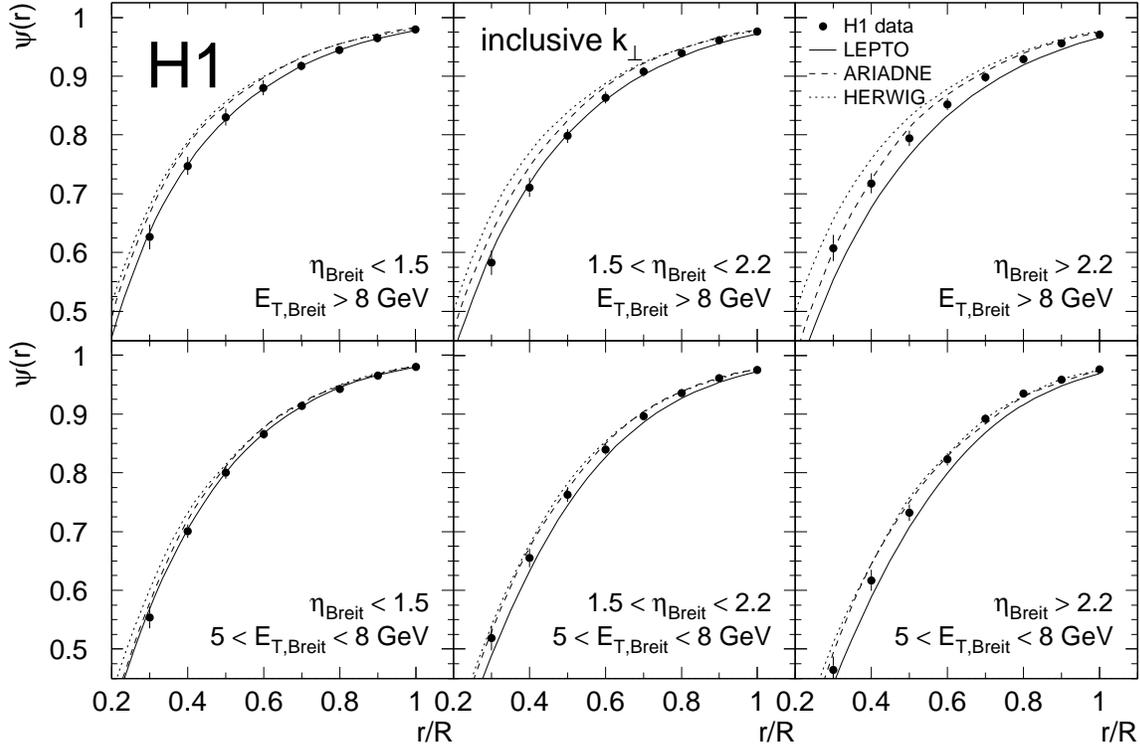,width=15.5cm}
\end{center}
\caption{The jet shapes for the inclusive $k_\perp$ algorithm.
The data are shown as a function of the transverse jet energy and the jet
pseudo-rapidity in the Breit frame (positive pseudo-rapidities
are towards the proton direction).
The results are compared to predictions of QCD models.
\label{fig:kt_shape_model}}
\end{figure}

The predictions of different QCD models are compared in
Fig.~\ref{fig:kt_shape_model} to the jet shapes measured 
for the $k_\perp$ algorithm.
The models LEPTO, ARIADNE and HERWIG all show \etbreit ~and 
\etabreit ~dependences similar to that seen in the data.
LEPTO gives the best description of the measured shapes for
$\etabreit < 2.2$ while at $\etabreit > 2.2$ the predicted 
jet shapes are too broad.
A reasonable description is also obtained by the ARIADNE model
except for jets at smaller pseudo-rapidities where the jet shapes
have the tendency to be too narrow.
For the HERWIG model the jet shapes are narrower than those
in the data in all \etbreit ~and \etabreit ~regions.
The same observations as above are made when comparing these 
QCD models with the subjet multiplicities and with the jet shapes 
for the cone algorithm (not shown here).

\begin{figure}
\begin{center}
\epsfig{file=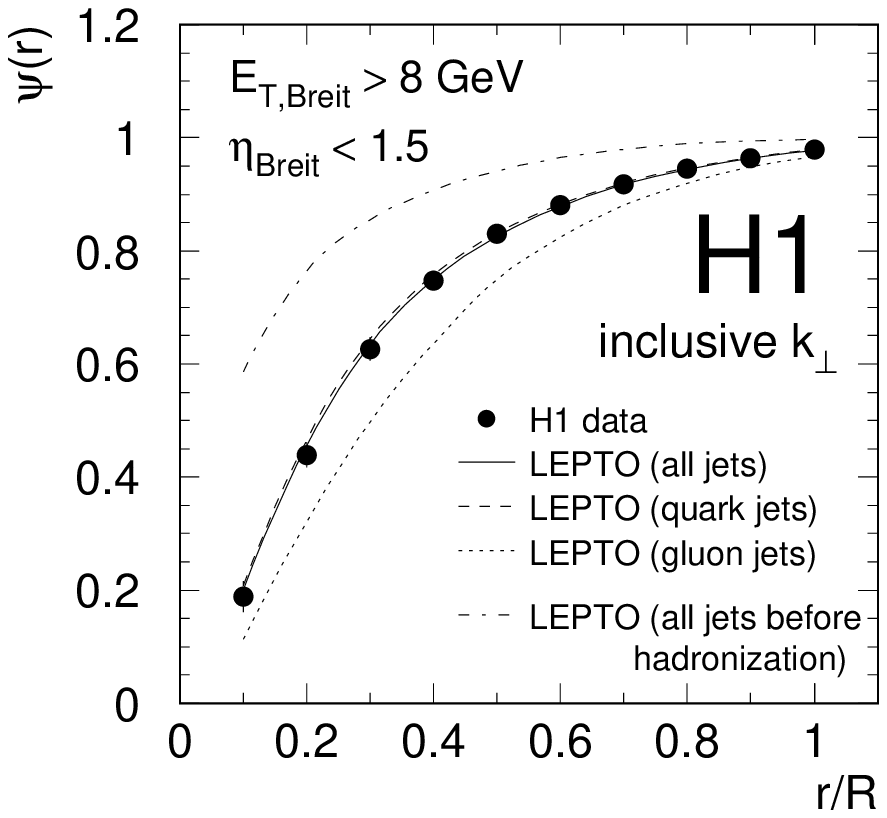,width=7.5cm}
\hskip7mm
\epsfig{file=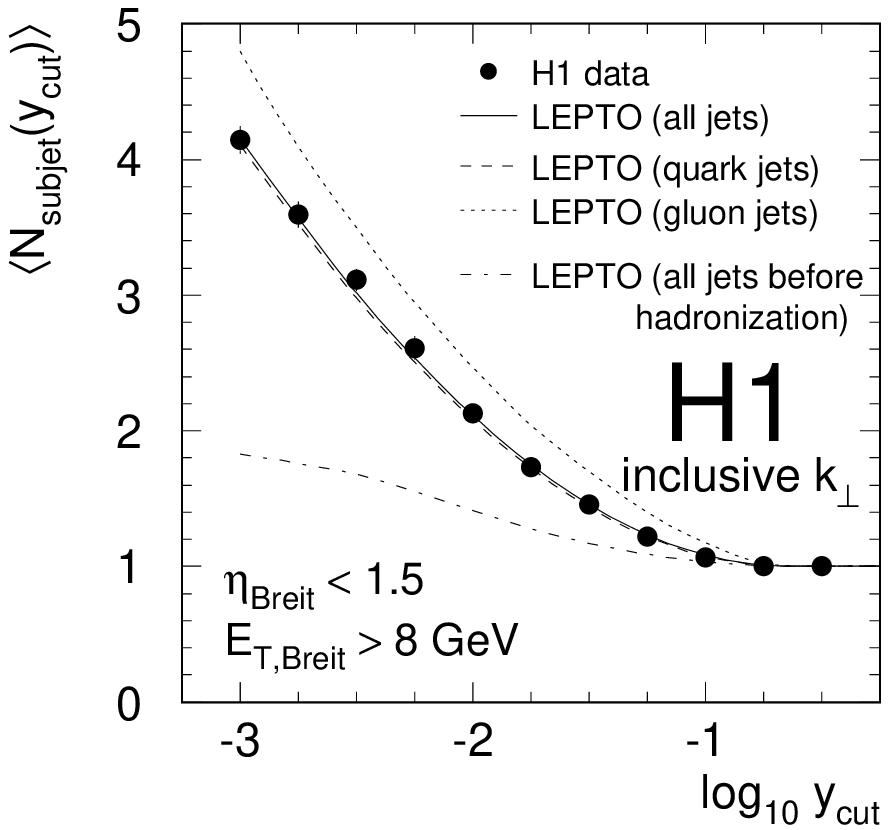,width=7.5cm}
\end{center}
\caption{
Model predictions of the internal structure of quark and gluon jets
for the inclusive $k_\perp$ algorithm by the LEPTO parton shower model.
The jet shapes (left) and the subjet multiplicities (right)
are shown separately for quark and gluon induced jets with 
$\etbreit > 8\,{\rm GeV}$ and $\eta_{\rm Breit} < 1.5$,
together with the sum of both and the comparison to the H1 measurement.
The distributions of the observables before hadronization are also shown.
\label{fig:models}}
\end{figure}

In QCD models the evolution of a jet is described by perturbative
contributions (radiation of partons) and non-perturbative contributions
(hadronization).
Studies based on the LEPTO and HERWIG parton shower models show 
that all observables studied in this analysis are strongly influenced 
by hadronization.
This process has the largest impact on the jet broadening in our
kinematic region (Fig.~\ref{fig:models}).
Basic characteristics of the perturbative contributions are 
however still visible after hadronization.
The model prediction suggests that the large difference between 
quark and gluon-initiated jets
before hadronization survives the hadronization process.
This especially applies to jets with large transverse energies 
~\cite{sonne}.

Fig.~\ref{fig:models} shows the jet shapes and the 
subjet multiplicities as predicted by the LEPTO
parton shower model for the $k_\perp$ algorithm, separately for
quark and gluon jets at $\etbreit > 8\,{\rm GeV}$ and $\etabreit < 1.5$.
Gluon jets are broader than quark jets.
The same prediction is obtained by the HERWIG parton shower model.
Although the jets in HERWIG are slightly narrower, the differences 
between gluon and quark jets are equally large.
In the phase space considered here, LEPTO and HERWIG (in agreement with 
next-to-leading order calculations) predict a fraction of approximately 
$80 \%$ photon-gluon fusion events with two quarks in the partonic final
state.
The jet samples of these models are therefore dominated by quark jets.
Both model predictions for the jet shapes and the subjet
multiplicities therefore mainly reflect the properties of the 
quark jets as can be seen in Fig.~\ref{fig:models}.
These predictions give a reasonable description of the data.
Thus, we conclude, that the jets we observe are consistent with 
being mainly initiated by quarks.

\section{Summary} 

Measurements of internal jet structure in dijet events
in deep-inelastic scattering in the kinematic domain 
$10 < Q^2 \lesssim 120\, \GeVsq$ and 
$2 \cdot 10^{-4} \lesssim x_{\rm Bj} \lesssim 8 \cdot 10^{-3}$
have been presented.
Jet shapes and subjet multiplicities have been studied
for jets of transverse energies $\etbreit > 5\,{\rm GeV}$ defined by 
$k_\perp$ and cone jet algorithms in the Breit frame.

The radial dependence of the jet shape and the dependence
of the average number of subjets on the subjet resolution parameter
$y_{\rm cut}$ are both sensitive to different aspects of jet broadening.
For both observables a dependence of the jet broadness on the 
transverse energy \etbreit ~and on the pseudo-rapidity in the 
Breit frame \etabreit ~is seen.
With increasing \etbreit ~jets are narrower. 
Jets of the same \etbreit ~become broader towards the proton direction.
This effect is more pronounced at lower \etbreit.

At lower \etbreit ~jets defined by the $k_\perp$ algorithm are 
more collimated than jets defined by the cone algorithm,
while at higher \etbreit ~both algorithms produce very similar jets.

The QCD models LEPTO, ARIADNE and HERWIG roughly reproduce the dependence 
of the jet shape and the subjet multiplicities on \etbreit ~and 
\etabreit ~as seen in the data.
LEPTO has a tendency to produce broader jets in the
proton direction than measured.
HERWIG and ARIADNE produce jets which are too collimated
especially at higher transverse energies. 
We have reported earlier that in the same kinematic domain the 
predicted jet rates from LEPTO and HERWIG are about a factor of 
two below the data~\cite{h1r2cone}.
Since these models are able to reproduce the internal jet structure,
this failure must be largely connected to an inadequate modeling
of the underlying hard partonic subprocess. 

According to the parton shower models LEPTO
and HERWIG, quark and gluon initiated jets differ both at the parton 
and at the hadron level.
Both models predict that the jet sample is dominated by quark
initiated jets.
Since these models describe our data, we conclude that the observed
jet structures are compatible with those of quark initiated jets.

\section{Acknowledgments}
We are grateful to the HERA machine group whose outstanding efforts
have made and continue to make this experiment possible.
We thank the engineers and technicians for their work in constructing and now 
maintaining the H1 detector, our funding agencies for financial support,
the DESY technical stuff for continual assistance, and the DESY directorate
for the hospitality which they extend to the non-DESY members of the
collaboration.

\bibliography{h1_jetstructure}

\newpage


\end{document}